\documentclass[11pt]{article}
\usepackage{graphicx}
\usepackage[margin=1.4in]{geometry}
\usepackage{amsfonts}
\usepackage{amsmath}
\usepackage[english]{babel}
\usepackage{hyperref}
\usepackage{color}
\hypersetup{colorlinks=true,citecolor=blue,linkcolor=blue,filecolor=blue,urlcolor=blue}
\usepackage{tikz}
\usepackage{framed}
\usepackage{setspace}
\usepackage{nicefrac}

\newcommand{\bq}{\begin{quotation}}
\newcommand{\eq}{\end{quotation}}
\newcommand{\be}{\begin{equation}}
\newcommand{\ee}{\end{equation}}
\newcommand{\bea}{\begin{eqnarray}}
\newcommand{\eea}{\end{eqnarray}}
\def\tr{{\rm tr}\,}

\title{
Quantum mechanics? It's all fun and games until \\ someone loses an $i$.\footnote{This paper is dedicated to Prof.\ Gopal Rao upon his promotion to Distinguished Professor Emeritus status. Prof.\ Rao once remarked to one of us (CAF) that being at UMass Boston---rather than say at Harvard or Yale---allowed his career to excel. UMass Boston brought him academic freedoms he could not find elsewhere: He worked on whatever he pleased, without pressure for high-dollar funding or worries over journals' impact factors. We offer this paper to Prof.\ Rao in his own spirit. True science is founded upon the freedom to become fascinated by a simple pebble on a beach, whether it be polished or not, or whether it have any value at the local market.}}

\author{Christopher A. Fuchs, Maxim Olshanii, and Matthew B. Weiss \medskip
\\
\small Department of Physics, University of Massachusetts Boston
\\
\small 100 Morrissey Boulevard, Boston MA 02125, \ USA}

\date{21 July 2022}

\begin{document}

\maketitle

\begin{abstract}
QBism regards quantum mechanics as an addition to probability theory. The addition provides an extra normative rule for decision-making agents concerned with gambling across experimental contexts, somewhat in analogy to the double-slit experiment. This establishes the meaning of the Born Rule from a QBist perspective. Moreover it suggests that the best way to formulate the Born Rule for foundational discussions is with respect to an informationally complete reference device. Recent work~\cite{DeBrota2020a} has demonstrated that reference devices employing symmetric informationally complete POVMs (or SICs) achieve a minimal \emph{quantumness}: They witness the irreducible difference between classical and quantum. In this paper, we attempt to answer the analogous question for real-vector-space quantum theory. While standard quantum mechanics seems to allow SICs to exist in all finite dimensions, in the case of quantum theory over the real numbers it is known that SICs do \emph{not} exist in most dimensions. We therefore attempt to identify the optimal reference device in the first real dimension without a SIC (i.e., $d=4$) in hopes of better understanding the essential role of complex numbers in quantum mechanics. In contrast to their complex counterparts, the expressions that result in a QBist understanding of real-vector-space quantum theory are surprisingly complex.
\end{abstract}

\section{Introduction}

Since the days of Heisenberg, Born, Jordan, Dirac, and Schr\"odinger in the mid-1920s, physicists have used the theory of quantum mechanics as it was taught to them.  But why just that theory and not some other?  The debate is ongoing to this day, and there is still plenty to learn.  One technique for better understanding why we use the formalism that we do is to consider ``foil theories'' in which some aspect of our usual quantum mechanics is either relaxed or restricted~\cite{Chiribella2015}. For example, one can consider a quantum-like theory without imaginary numbers and try to see what ``goes wrong.'' This is a conceptual game with a long and distinguished history~\cite{Jordan1932,Jordan1934,Birkhoff1936,Stueckelberg1960,Stueckelberg1961,Wootters1980,Lahti1987,Wheeler1988,Aaronson2004,Baez2012,Aleksandrova2013,Wootters2013}.  In this setting, probabilities are still given by the squares of amplitudes, but now amplitudes are drawn from vectors in a real vector space, where the phases are simply $\pm 1$. Similarly, density matrices---positive semi-definite Hermitian matrices of unit trace---are replaced by their real counterparts, positive semi-definite symmetric matrices, and the unitary matrices furnishing time evolution are replaced by real orthogonal matrices (i.e., simple rotations). The hope is that by contemplating such a theory, one can begin to ``see around'' standard quantum mechanics and start to understand what is genuinely unique about it.

A case in point has to do with the QBist interpretation of quantum mechanics~\cite{Fuchs2010a,Fuchs2014a,Fuchs2017a,DeBrota2018}.\footnote{From here out, we reserve the term ``quantum mechanics'' for normal complex-vector-space quantum theory. Whereas, when speak of ``a quantum theory,'' this generally will include the possibility that it could also be a quantum-like foil theory.}  QBism stresses that it is possible to express any quantum-like foil theory (over any number field) purely in terms of measurement-outcome probabilities, without ever referencing state vectors, amplitudes, or operators~\cite{Fuchs2011b}. From this point of view, the Born Rule is regarded as a physically motivated {\it addition\/} to the usual Law of Total Probability (LTP)~\cite{Fuchs2013a}. It is an addition useful for transferring one's expectations from one experimental situation to another, a situation the LTP has no jurisdiction over. The exact expression the Born Rule takes, however, depends on one's choice of a ``reference device''~\cite{DeBrota2020b}. 

One might wonder, then, which reference devices minimize the difference between the Law of Total Probability and the Born Rule---in other words, which reference devices witness the irreducible difference between classical and quantum uses of probability, by some measure of ``quantumness.'' In the case of quantum mechanics, the answer has been provided by DeBrota, Fuchs, and Stacey~\cite{DeBrota2020a}: the optimal reference device is one which employs a {\it symmetric informationally complete\/} measurement. Such measurements 
are often called simply SICs (pronounced ``seeks'') for short.

More formally, suppose one has a $d$-level quantum system, a {\sl qudit}.  A set of $d^2$ state vectors $|\psi_k\rangle$ satisfying,
\be
|\langle\psi_j|\psi_k\rangle|^2=\frac{1}{d+1}\quad \forall j\ne k\;.
\label{Oneness}
\ee
is known as symmetric informationally complete, and when the projectors onto these vectors are rescaled to
\be
R_k = \frac{1}{d}|\psi_k\rangle\langle\psi_k|\,,\quad k=1,2,\ldots d^2\,,
\label{Twoness}
\ee
the collection represents the possible outcomes of a reference measurement on the qudit.  What licences the designation of such a device as a reference-measurement is that the operators $R_k$ can be proven to be linearly independent, and since there are $d^2$ of them they will span the space of Hermitian operators.  Thus they form a basis for that space.  On the other hand the symmetry of the set is apparent in Eq.~(\ref{Oneness}).  Since one can think of the projectors onto the vectors as specifying lines in a $d$-dimensional space, these structures are also known as maximal sets of complex equiangular lines.\footnote{After 23 years of research, it remains an open question whether SICs in fact exist in all complex dimensions.  See Ref.~\cite{Fuchs2017b} for a review.  However, that does not mean the SICs cannot already be a playground for better understanding physics.  Currently, exact constructions of SICs can be found in all dimensions $d\le 53$ and for 72 specific dimensions beyond that, going all the way out to $d=39,604$~\cite{Appleby2021}.  Furthermore, there is high-precision numerical evidence for all dimensions $d=54$ to 193.  See Refs.~\cite{Grassl2021,Grassl2022}. There is a general belief in the community that SICs exist in all finite dimensions, but until a proof of such, it is only an educated guess.}

\begin{table}
$$
\begin{tabular}{|c|c|c|}
\hline
 $d$ & Dimension of Operator Space & $N_{\rm max}$ \\ [0.5ex]
 \hline\hline
  2 & 3 & 3 \\
  3 & 6 & 6 \\
  4 & 10 & 6 \\
  5 & 15 & 10 \\
  6 & 21 & 16 \\
  7 & 28 & 28 \\
  8 & 36 & 28 \\
  9 & 45 & 28 \\
  \vdots & \vdots & \vdots \\
  14 & 105 & 28 \\
  15 & 120 & 36 \\
  16 & 136 & 40 \\
  \vdots & \vdots & \vdots \\
  23 & 276 & 276 \\
  24 & 300 & 276 \\
\hline
\end{tabular}
$$
\label{JimmySmith}
\caption{The number of elements required for the analogue of a SIC in a real vector space versus the actual maximum number of equiangular lines~\cite{Sloan2015} in that dimension. One sees that $N_{\rm max}$ achieves the upper bound only in dimensions $d=2,3,7$ and $23$.  It is not known whether $N_{\rm max}$ achieves the upper bound in any further dimensions.}
\end{table}

In this paper, we consider the analogous question in the setting of real-vector-space quantum theory, and offer some preliminary results. One might think that the analogue of a SIC in this setting would correspond to a maximal set of \emph{real} equiangular lines.  However, there is a catch.  A minimal informationally complete measurement for a $d$-level system in real-vector-space quantum theory (a RIC) requires $\frac{1}{2}d(d+1)$ POVM elements in order to match the dimension of the symmetric matrices.  But it is known that $\frac{1}{2}d(d+1)$ only provides an {\it upper bound\/} on the actual maximal number $N_{\rm max}$ of equiangular lines---a bound that is sometimes achieved, but mostly not~\cite{Lemmens1973}. 

As it turns out, the bound is tight in $d=2$ and $3$: in $d=2$, $N_{\rm max}=3$ and in $d=3$, $N_{\rm max}=6$. Therefore, in $d=2$ and $3$,  SIC-POVMs exist. However as stated, the bound is \emph{not} tight in most real dimensions, as Table \ref{JimmySmith} attests. 

What is of interest to us in this paper is a $d=4$ system\footnote{This is mathematically equivalent to the case of two {\it rebits}~\cite{Caves2001}.  Viewing it that way, i.e., as a bipartite system, there is a significant literature on its ``broken'' notion of a tensor product and the similarly problematic concept of entanglement that comes with it~\cite{Araki1980,Wootters1990,Caves2002b,Wootters2012,Hardy2012,Renou2021,Li2022,Chen2022}.  Herein however, we will always think of $d=4$ as associated with {\it a single system}, as for instance with a four-level atom where there is no natural notion of two subsystems. A consequence of this is that the ``broken thing'' we demonstrate in this paper will be of quite a distinct character from the ones to do with entanglement.}, one with the lowest dimension for which $N_{\rm max}\ne \frac{1}{2}d(d+1)$. There, the maximum number of equiangular lines is $6$, but one requires $10$ elements to span the space of real density matrices.

The broadest question on our minds is what might a QBist stand to learn about normal quantum mechanics by studying this case?  Particularly, what is the stand-in for the result of Ref.~\cite{DeBrota2020a} mentioned earlier?  What reference devices in real-vector-space quantum theory witness the irreducible difference between classical and quantum uses of probability theory?  Moreover, when one uses that optimal device to express the Born Rule, how does the expression compare to the one found in normal quantum mechanics?

The main message of this paper is that in normal quantum mechanics the Born Rule remains relatively elegant in appearance when expressed in irreducible QBist form, whereas in real-vector space quantum theory, the irreducible form is genuinely ugly by any aesthetic measure. In fact we must admit that when we first embarked on this project, we did expect the expression to be a little ugly (that was our desired result). However, we were quite unprepared for the magnitude of the ugliness we ultimately found.  (For a preview, see Eq.~(\ref{HankSnow}).)  Moreover, in contrast to the quantum mechanical case, the irreducible form appears not to be unique, having a delicate dependence on the norm used for defining it.  So whereas the quantum mechanical concept is a robust one, in the real-vector-space case, the notion of an irreducible form appears to be flawed from the outset.  To a QBist nose, there certainly seems to be a lesson in this.  \v{C}aslav Brukner likes to ask \cite{BruknerForever} what is so special about regular quantum mechanics for the QBist, since one can ostensibly give a QBist interpretation to any generalized probabilistic theory (GPT)~\cite{Plavala2021}?  Maybe the answer is this:  For normal quantum mechanics, the various QBist-inspired developments of the formalism seem to fit the theory like a glove. But if the real-vector-space theory exhibits the more common behavior among GPTs, and it is indeed so ill-fitting, one could question whether it is so fruitful to think in QBist terms for that theory in the first place. Like the Bohmian rewriting of quantum mechanics, it can be done, but at what cost?

The plan of the remainder of our paper is as follows.  In Section~\ref{StarTrekDay!}, we review how DeBrota, Stacey, and Fuchs~\cite{DeBrota2020a} set up the problem in regular complex-vector-space quantum mechanics and exhibit the result found there.  In Section~\ref{PuddinTame}, we recount our initial stabs in the dark toward an optimal RIC-POVM reference device, beginning with certain known symmetrical polytopes and ending with sampling from the space of RICs. In Section~\ref{AskMeAgain}, we discuss the initial results of constrained optimization of the quantumness over unbiased rank-1 RICs following the parallel-update rule. In Section~\ref{I'llTellYouTheSame}, we consider biased RICs and uncover an intriguing parametric structure that offers a different optimal RIC for each choice of $p$-norm. Then in Section~\ref{BringingHome}, we realize that allowing a distinct post-measurement offers an opportunity for even lower quantumness, for which we are yet to have an analytic expression.  Finally, in Section~\ref{Lumpitude} we conclude with some remarks on the significance of this work and  further directions that might be taken.

\section{Review of the Quantum Mechanical Case}
\label{StarTrekDay!}

QBism begins with the observation that, instead of working with density matrices and measurement operators for all one's quantum mechanical calculations, one can work just as well (if perhaps inconveniently) with probabilities and conditional probabilities for the outcomes of a fixed {\it reference device}. This is singularly important to the interpretation, for if it were not true, one might be tempted (as many philosophers of physics are~\cite{Brown2019}) to view quantum states as something more substantial than personal degrees of belief.  To see how the translation works, first recall some concepts from quantum information theory.

Throughout we will restrict ourselves to finite dimensional quantum systems. For this section, let $\mathcal{H}_d$ be a $d$-dimensional complex Hilbert space, and
let $\{E_j\}$ be a set of $N$ positive semidefinite operators whose elements sum to the identity operator:
\begin{equation}
\sum_{j=1}^N E_j = I.
\end{equation}
Such sets are called \emph{positive-operator-valued measures} (POVMs) and represent the most general measurements allowed in quantum mechanics, where $N$ is any nonnegative integer. The elements of the set stand for the $N$ possible {\it outcomes\/} of the measurement~\cite{Nielsen2010}.\footnote{Note how this differs from the treatment of measurement one finds in textbooks from the pre-quantum-information era. There a measurement is associated with a {\it single\/} Hermitian operator, and the {\it outcomes} correspond to the operator's eigenvalues.  Here, however {\it the operators are the outcomes}, and particularly the number of outcomes $N$ can exceed the dimensionality of the underlying Hilbert space.}  A POVM is said to be
\emph{informationally complete} (IC) if the $E_j$ span
the space of Hermitian operators on $\mathcal{H}_d$, and
an IC-POVM is said to be \emph{minimal} if it contains exactly $N=d^2$
elements---i.e., it forms a {\it basis\/} for the space.  For brevity, we will call a minimal informationally complete
POVM a MIC (pronounced ``meek''), and if all the elements of $\{E_j\}$ are rank-1, we will call it a rank-1 MIC.

The standard procedure in quantum mechanics for generating probabilities starts with an observer or agent, say Alice, assigning a quantum state $\rho$ to a system. When she plans to measure the system, she represents the outcomes of her measurement with a POVM $\{E_j\}$.  Assigning $\rho$ implies that Alice assigns the Born Rule probabilities
\be
Q(E_j)=\tr E_j \rho
\ee
to the measurement's outcomes. In this way, any quantum state $\rho$ may be thought of as a catalog of probabilities for all possible measurements. However one does not have to consider all possible measurements to completely specify $\rho$.  Because MICs form bases for the space of operators, $\rho$ is uniquely specified by the agent's expectations for the outcomes of any single MIC:  Indeed,  $Q(E_j)$ represents the Hilbert-Schmidt inner product between $\rho$ and  $E_j$, and if one knows $\rho$'s projections onto a basis, then one knows $\rho$ itself. Thus with respect to any MIC, any quantum state, pure or mixed, is equivalent to a single probability distribution.

\begin{figure}
\begin{center}
    \includegraphics[width=4.5in]{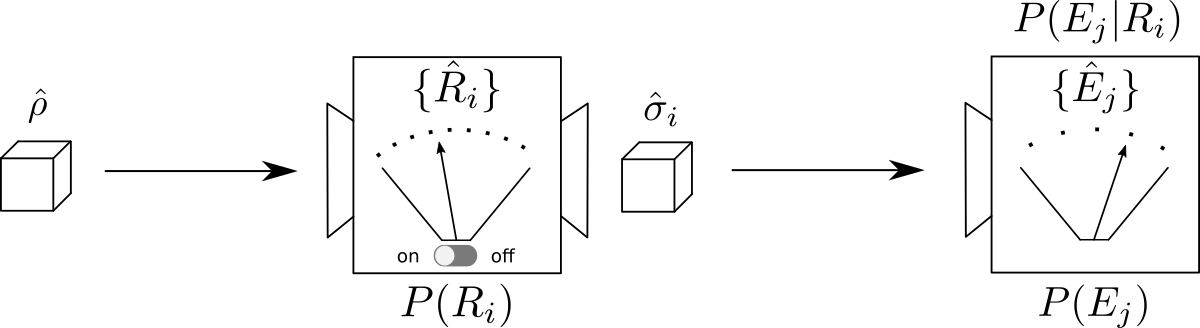}
\end{center}
\begin{center}
    \includegraphics[width=4.5in]{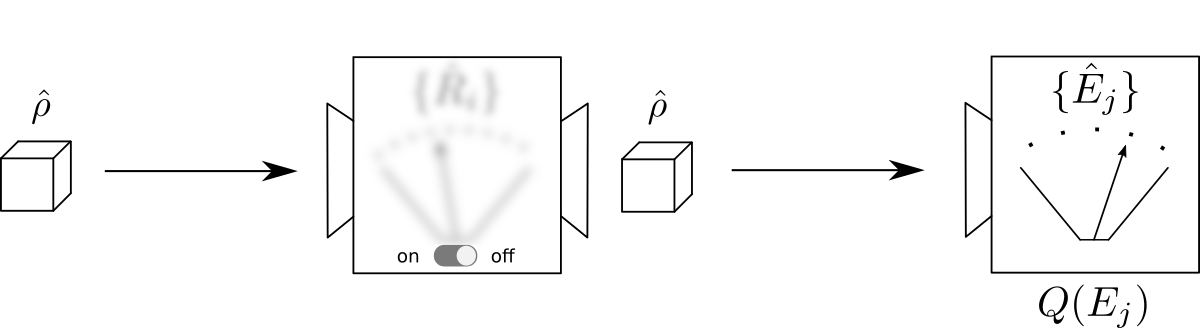}
\end{center}
\caption{Two distinct experiments.  In QBism, the Born Rule is not about either experiment individually, but rather about the connections between their probabilities.  In the top experiment, the reference device is turned on so that there are three probabilities in its telling ($P(R_i), P(E_j|R_i), P(E_j)$): they must satisfy the Law of Total Probability, Eq.~(\ref{ltp}). However, in the bottom experiment the reference device is turned off---there is only one probability in its story ($Q(E_j)$). The Born Rule is the narrative glue that ties the two stories together.}
\label{Metzenbaum}
\end{figure}

One can further eliminate the need to use the operators $\rho$ and $E_j$ in the Born Rule by reexpressing it as a relation between the agent's expectations in two distinct experiments (see Figure~\ref{Metzenbaum}). Suppose Alice has a preferred reference device consisting of a MIC $\{R_i\}$ followed by a post-measurement preparation of the quantum system: If the MIC obtains outcome $R_i$, a new state $\sigma_i$ will be ascribed to the system.  We will require that the $\sigma_i$ be drawn from a linearly independent set, but otherwise the set may be arbitrary. The reason for the linear independence is that we want the inner products $\tr E_j\sigma_i$ to uniquely characterize the operators $E_j$. Let $P(R_i)$ be the agent's probabilities for the measurement $\{R_i\}$ and
\be
P(E_j|R_i)=\tr E_j \sigma_i
\ee
be her conditional probabilities for a subsequent measurement of $\{E_j\}$ after obtaining outcome $R_i$. What consistency requirement among $Q(E_j)$, $P(R_i)$, and $P(E_j|R_i)$ does quantum physics entail?

Using the fact that $\{\sigma_i\}$ is a basis, we may write
\begin{equation}
    \rho=\sum_j\alpha_j\sigma_j\;,
\end{equation}
for some set of real coefficients $\alpha_j$.  The probability of outcome $R_i$ is then
\begin{equation}
    P(R_i)=\sum_j\alpha_j\,\tr R_i\sigma_j=\sum_j\,\big[\Phi^{-1}\big]_{ij}\,\alpha_j\;,
\end{equation}
where we have defined the ``Born matrix'' $\Phi$ through its inverse,
\begin{equation}\label{phiinv}
    \big[\Phi^{-1}\big]_{ij}:=\tr R_i\sigma_j=r_i\tr\rho_i\sigma_j\;,
\end{equation}
for $\rho_i:=R_i/r_i$ and $r_i:=\tr R_i$. The invertibility of $\Phi$ is assured by the linear independence of the MIC and post-measurement sets. This implies that the coefficients of $\rho$ in the $\sigma_i$ basis may be written as the multiplication of the $\Phi$ matrix onto the vector of probabilities,
\begin{equation}\label{rhoinbasis}
    \rho=\sum_i\left[\sum_k[\Phi]_{ik}P(R_k)\right]\! \sigma_i\;.
\end{equation}
Now, the probability $Q(E_j)$ can finally be revealed by another application of the Born Rule, which becomes
\begin{equation}\label{ltpanalogindices}
                Q(E_j)=\sum_{i=1}^{d^2}\left[\sum_{k=1}^{d^2}[\Phi]_{ik}P(R_k)\right]\! P(E_j|R_i)\;.
\end{equation}
In short, the Born Rule is purely about the relation between the probabilities in the two experiments.

In a more compact matrix notation, our result becomes particularly evocative.  Let $P(R)$ and $Q(E)$ denote vectors whose components are $P(R_i)$ and $Q(E_j)$ respectively, and $P(E|R)$ denote an appropriately sized stochastic matrix. Then, Eq.~(\ref{ltpanalogindices}) becomes
\be
Q(E) = P(E|R)\Phi P(R)\;.
\label{ltpanalog}
\ee
Note how similar, yet different, this is to the Law of Total Probability, which only
relates the probabilities in the top experiment in Figure~\ref{Metzenbaum} together:
\be
P(E) = P(E|R)P(R)\;.
\label{ltp}
\ee
The only difference between the right-hand sides of Eqs.~(\ref{ltpanalog}) and (\ref{ltp}) is that in the
first, the two terms are separated by $\Phi$ and in the second they are implicitly separated by the identity $I$.\footnote{Mathematical expressions for the Born rule with forms similar to Eq.~(\ref{ltpanalog}) go back at least to the work of Lucien Hardy in 2001.  See Ref.~\cite{Hardy2001}.} In
fact, before knowing any quantum mechanics, one's intuition might have been that $Q(E)$
ought to just be $P(E)$. But that is an intuition drawn from classical physics, where the
role of experiment in shaping reality is thought to be ultimately eliminable.

This point raises an interesting mathematical question for the QBist. Depending
upon which reference device the agent chooses for their QBist representation, Eq.~(\ref{ltpanalog})
can be made to look more or less like the classical LTP. If one could find a reference
device so that $\Phi = I$, then one would have the LTP identically, and classical intuition
would be validated after all. But there is no such reference device~\cite{Fuchs2002}. So, how close
can $\Phi$ be made to look like the identity? The answer to this question would establish
an important fact about quantum mechanics: It would signal the irreducible difference
between the Born Rule and the classical intuition that would seek to set $Q(E) = P(E)$
if not impeded.

Fuchs, DeBrota, and Stacey~\cite{DeBrota2020a} quantified this question by introducing a class of
distance functions (or quantumness measures) based on unitarily invariant norms
\be
d(I, \Phi ) = \| I-\Phi \|\, .
\label{Ding-ding-ding}
\ee
A unitarily invariant norm is a matrix norm for square matrices such that $\|UXV\|=\|X\|$ for any unitary matrices $U$ and $V$.  These norms form a significant class in matrix analysis~\cite{Horn1994} and include the Schatten $p$-norms
\begin{equation}
||X||_{p} = \left(\sum_{i} s_{i}^p	\right)^{\!\frac{1}{p}},
\end{equation}
among which are the trace norm, the Frobenius norm, and the operator norm when $p=1,2,$ and $\infty$ respectively, and the Ky Fan $k$-norms. Here the $s_i$ represent the singular values of $X$. The class of $\Phi$ matrices that achieve the minimal distance from the identity $I$ define the {\it irreducible quantumness\/} of the Born Rule.

To set ourselves up for expressing the irreducible quantumness, let us say a bit more about the SICs first.  A SIC is a MIC for which all the $R_i$ are rank-1 and
\begin{equation}
\tr R_i R_j = \frac{1}{d^2} \frac{d\delta_{ij} + 1}{d+1}\;.
\end{equation}
SICs have yet to be proven to exist in all finite dimensions $d$, but they are widely believed to~\cite{Fuchs2017b}, and have even been experimentally demonstrated in some low dimensions~\cite{Durt2008,Medendorp2011,Zhao2015}.  The \emph{SIC projectors} associated with a SIC are the pure states $\rho_i=dR_i$. When there is no chance of confusion, we will refer to the set of projectors as SICs as well. In the past, QBism has given special attention to the case of a reference device whose measurement is a SIC and whose post-measurement states are SIC projectors associated with the same SIC~\cite{Fuchs2011b,Fuchs2017a,Appleby2017b}, but in all cases previous to Ref.~\cite{DeBrota2020a}, it was essentially for aesthetic reasons.  In this special case we denote $\Phi$ by $\Phi_{\rm SIC}$ and note that Eq.~\eqref{ltpanalogindices} takes a particularly simple form
\begin{equation}
Q(E_j)=\sum_{i=1}^{d^2}\left[(d+1)P(R_i)-\frac{1}{d}\right]\! P(E_j|R_i)\;.
\label{urgleichung}
\end{equation}
In other words, the total action of $\Phi_{\rm SIC}$ is a component-wise affine transformation of the probability vector.  If one had to generalize away from the LTP, what could be a simpler modification of it?

Now for the result of Ref.~\cite{DeBrota2020a}.  It can be proven that for all the distance measures considered in Eq.~(\ref{Ding-ding-ding}) and for all reference devices,
\be
d\big(I, \Phi \big)\ge d\big(I, \Phi_{\rm SIC}\big)
\ee
with equality if and only if the reference device measures a SIC and outputs post-measurement states that are also elements of a SIC.
That is, $\Phi_{\rm SIC}$ is not only an aesthetic choice, but one that tells us something deep about the very structure of quantum mechanics.

However, as we have observed, a SIC generally does not exist in real-vector-space quantum theory.  What can that be telling us about the foil theory?  We will study this in detail in the remaining sections.  Going forward, it is worth noting some of the aspects particular to the SIC reference devices in quantum mechanics:
\begin{enumerate}
\item {\bf Unbiasedness:} The trace of each POVM element $R_i$ in the reference device is the same---i.e., it is equally weighted. If the quantum state is $\rho=I/d$, the outcomes of $\{R_i\}$ will thus be equally probable.  In a general reference device, the weights might be different from each other, representing POVMs for which some outcomes are intrinsically more likely than others.
\item {\bf Rank-1:} Each element can be written in the form $R_{i} = \alpha_i |\psi_{i}\rangle\langle \psi_{i}|$ for $\alpha_i>0$. Thus we can also consider a SIC-POVM to correspond to frame theory's notion of a {\it tight vector frame}~\cite{Waldron2018} in ${\cal H}_d$.  More generally, a MIC need not have rank-1 elements.
\item {\bf Equiangularity:} $\tr R_i R_j = c$ when $i\ne j$.  This of course is part of the defining condition for a SIC, but it can also be achieved by non-rank-1 MICs. Either way, it is already a very restrictive condition on a MIC.
\item {\bf Robust Minimality:} Using SICs for both the POVM elements and the post-measurement states of a reference device minimizes the quantumness $||I-\Phi||$ with respect to {\it any\/} unitarily invariant norm.  One can imagine a world where that might not have been the case---where each norm would need a separate treatment---but that is not the case with quantum mechanics.
\item {\bf Parallel Update:} In the case where the reference POVM $\{R_i\}$ is a SIC, the post-measurement states $\sigma_i$ can be chosen to be drawn from the {\it same\/} SIC without loss of generality. However, there is nothing in the definition of irreducible quantumness that would make that property a priori obvious.
\end{enumerate}

As we now begin to identify the reference devices for achieving the irreducible representation of the Born Rule for $d=4$ real-vector-space quantum theory, we shall see which of these properties have to be compromised to get there.

\section{Initial Considerations}
\label{PuddinTame}

Before going forward, let us review a key theorem from Ref.\ \cite{DeBrota2020b} upon which much of our analysis is based.  Consider a set of $N$ normalized vectors $|\phi_i\rangle$ in $\mathcal{H}_d$ with weights $0\leq e_i\leq 1$. Then $E_i=e_i|\phi_i\rangle\langle\phi_i|$ forms a rank-$1$ POVM if and only if the ``little Gram matrix'' $g$ defined by
\be
[g]_{ij}=\sqrt{e_ie_j}\langle\phi_i|\phi_j\rangle
\ee
is a rank-$d$ projector.  Furthermore, defining the ``big Gram matrix'' $G$ by
\be
[G]_{ij}=\tr E_i E_j\;,
\ee
the set $\{E_i\}$ will be a rank-1 RIC if and only if $N=\frac{1}{2}d(d+1)$ and $\mbox{rank}\; G = N$.

Also in light of what follows, we note that if a SIC \emph{had} existed in $d=4$, assuming the parallel-update rule, its $\Phi$ matrix would have been
\begin{align}
    \Phi_{ij} &= \left(\frac{d+2}{2}\right)\delta_{ij} - \frac{1}{d+1}\\
    &= 3\delta_{ij} - \frac{1}{5}
\end{align}
Thus $I-\Phi$ would have one singular value of $0$ and $N-1=9$ singular values equal to $\frac{d}{2}=2$, leading to a $p$-quantumness of $ 2\times9^{\frac{1}{p}} $. In particular, when $p=2$, we obtain $6$.

Our initial hope was that the reference device achieving the irreducible representation for $d=4$ real-vector-space theory would be related to some long-known symmetric polytope already available in the literature. For example, the \emph{rectified 4-simplex} has an intimate connection with the famed \emph{Petersen graph}, containing 10 vertices and 15 edges. One can form a matrix whose rows and columns are labeled by the graph's vertices, whose elements are: $\frac{2}{5}$ along the diagonal, $-\frac{4}{15}$ if the two vertices are connected by an edge, and $\frac{1}{15}$ if not~\cite{Bachoc2009}. This $10 \times 10$ matrix is a rank-4 projector, and so corresponds to a rank-1 POVM which we shall call the Petersen RIC~\cite{Khanal2020}; the vectors $|\phi_i \rangle$ can be recovered from the little Gram matrix via a singular value decomposition. Since there are only two unique inner products between elements of the Petersen RIC, one could justly hope that its quantumness might be minimal. Assuming the parallel-update rule, the 2-quantumness (i.e., defined with respect to the Schatten 2-norm) of this reference device is $6 \sqrt{\frac{161}{5}} \approx 34.05$.

Next, we considered a RIC conjured from the so-called \emph{runcinated 5-cell} ~\cite{Coxeter1973,RunciFive}, a 4-polytope with 20 vertices, which come in antipodal pairs, picking out 10 lines in 4 dimensions. In fact, the vertices are root vectors of the simple Lie group\footnote{In fact, one can build a RIC in any dimension $d$ out of the root vectors of $A_d$. In dimension 2, one obtains a hexagon, whose three diagonals form an equiangular set---in other words, a SIC-POVM whose 2-quantumness is $\sqrt{2}$. In dimension 3, one obtains a RIC built from the cuboctahedron, whose 2-quantumness is $\sqrt{21} \approx 4.58$, which one can compare to the quantumness of the SIC derived from the 6 diagonals of the icosahedron, the latter being $\frac{3\sqrt{5}}{2} \approx 3.35$.} $A_4$. In dimension 4, again assuming the parallel-update rule, the $A_4$-RIC has 2-quantumness $2\sqrt{21} \approx 9.17$, kicking the Petersen RIC out of the water, and coming quite close to the 2-quantumness of the non-existent SIC. Even better, its POVM elements also have but two unique inner products between them.

\begin{figure}
\begin{center}
    \includegraphics[width=2in]{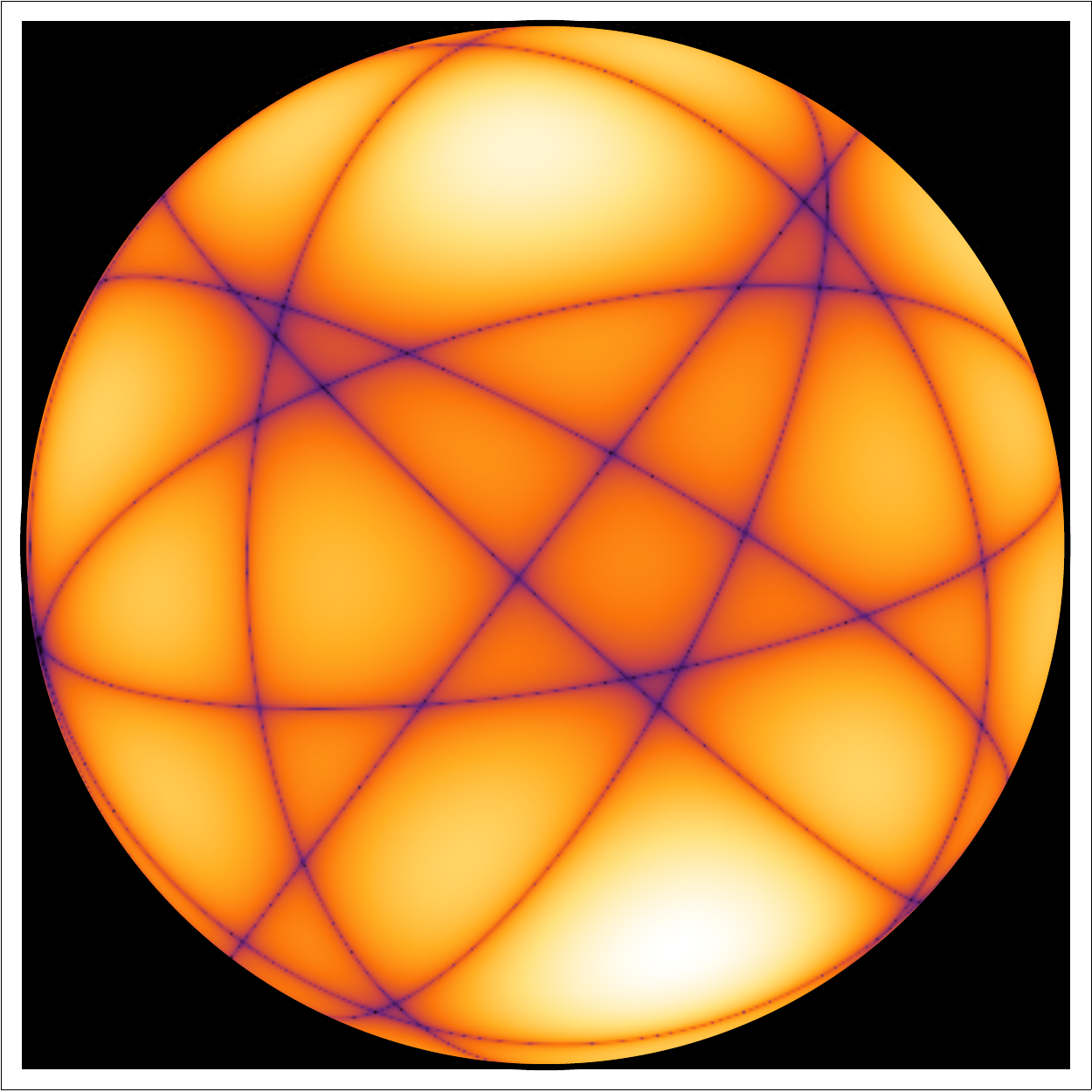}
    \hspace{1cm}\includegraphics[width=2in]{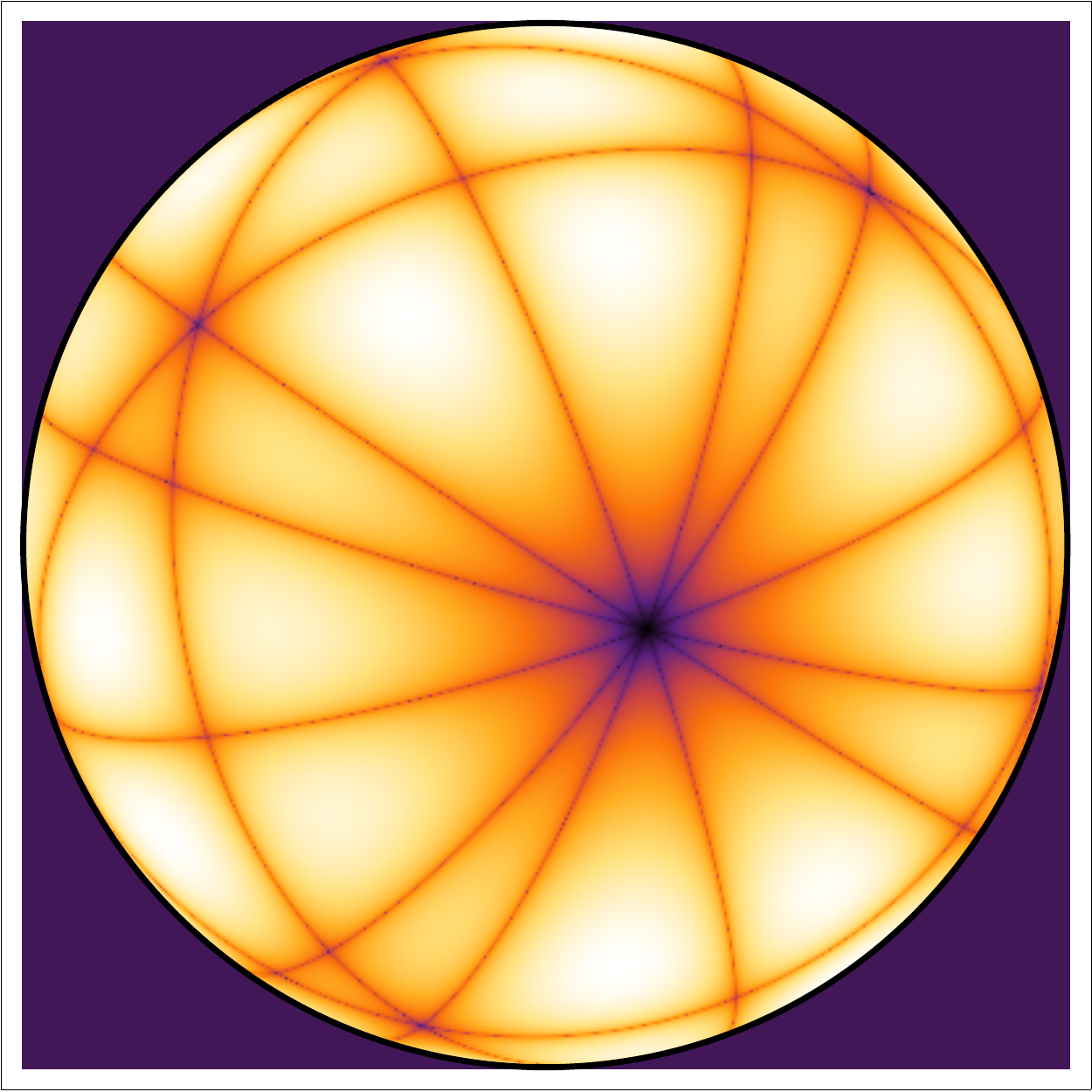}
\end{center}
\caption{On the left, a visualization of the Petersen RIC; and on the right, a visualization of the $A_4$-RIC.}
\label{PetersenAndA4}
\end{figure}

For insight into the structure of these four dimensional objects, one can proceed as follows to visualize them \cite{Olshanii2015}. Let
\begin{equation}
    \psi(|z\rangle) = \prod_{i=1}^{10}  \langle z | \phi_i \rangle
\end{equation}
where $|z\rangle = [z_1, z_2, z_3, z_4]$. Then pick a 3-sphere and a plane, e.g.
\begin{align}
    \langle z | z \rangle = R^2 &&  | z \rangle  \perp [0, 1, -1, -1],
\end{align}
from some choice of $R$. At the intersection of the 3-sphere and the plane, the function $\psi(|z\rangle)$ reduces to a function over the 2-sphere. Each vector $|\phi_i \rangle$ is represented by a great circle  on the sphere, and the angles between the circles faithfully represent the angles between the original four dimensional vectors. (See Figure~\ref{PetersenAndA4}.)

\begin{figure}
\begin{center}
    \includegraphics[width=3.5in]{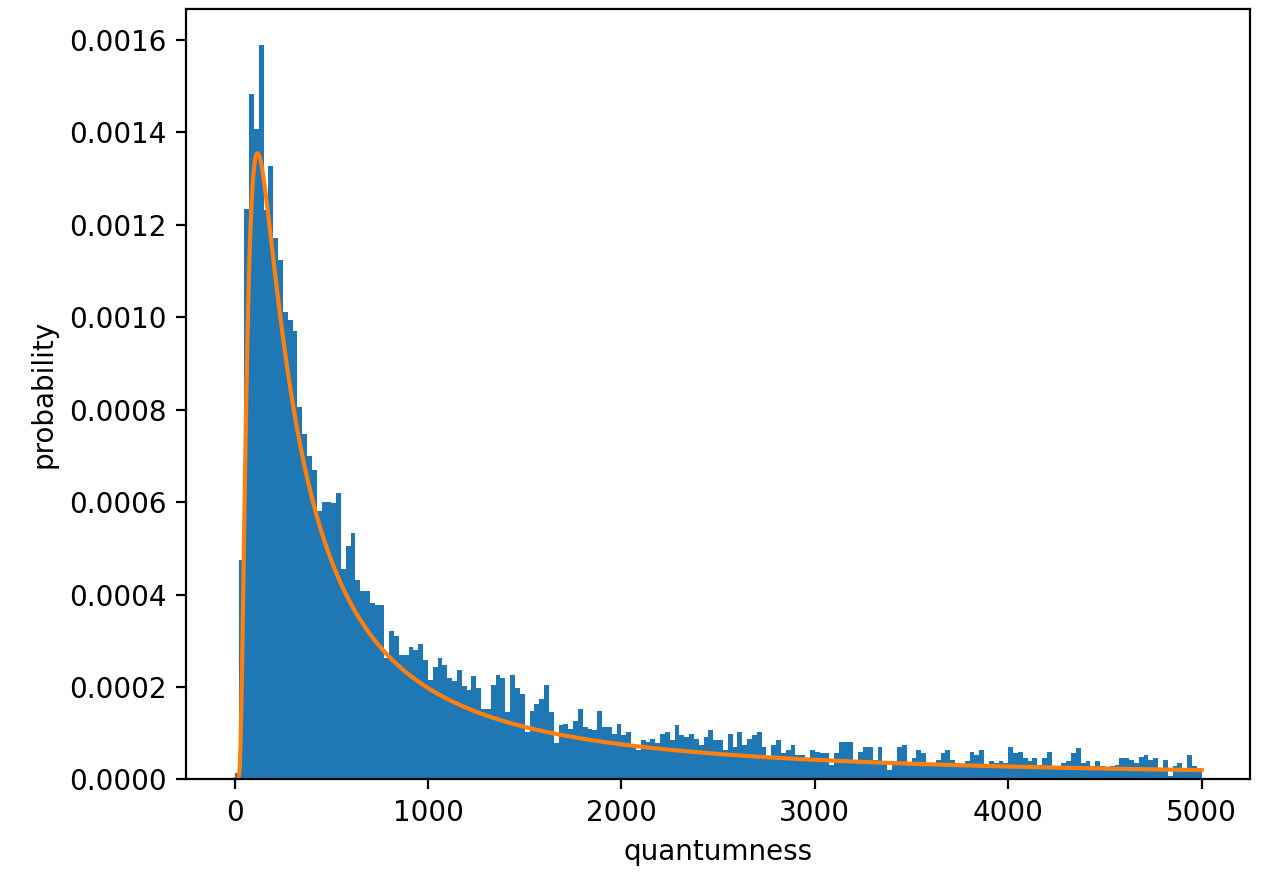}
\end{center}
\caption{Histogram of the 2-quantumness for 10,000 randomly sampled RICs.}
\label{Levy}
\end{figure}

Of course, guessing the answer can only take one so far. We thus began a series of numerical experiments in order to survey the terrain of RIC-POVMs, at first sampling at random from the landscape. For example, Figure~\ref{Levy} offers a histogram of the values of the 2-quantumness for $10,000$ random RICs fit to a L\'evy distribution, whose pdf is
\begin{equation}
    f(x) = \frac{1}{a\sqrt{2\pi \Big(\frac{x-b}{a}\Big)^{\!3}}}e^{-\frac{a}{2(x-b)}},
\end{equation}
with scale parameter $a \approx 341.31$ and shift parameter $b \approx 5.12$. It peaks at $\approx 120$ with a long tail thereafter. The lowest value found for the 2-quantumness was $\approx 16$, leaving the $A_4$-RIC unchallenged. Indeed, adding noise to the $A_4$-RIC's elements only ever increased its 2-quantumness.

We then considered randomly sampling from the space of unbiased rank-1 RICs using an alternating projection method: Beginning with a matrix of initial (row) vectors $|\phi_i\rangle$, we alternate between a) enforcing the POVM condition by taking the generalized polar decomposition $F = UP$ for $U$ an isometry and $P$ Hermitian, thereafter throwing away the Hermitian part; and b) normalizing the vectors---until we have an unbiased rank-1 RIC up to some desired tolerance. Moreover, we experimented with adding a third projection, knocking down the quantumness, whereby at each step one calculates the singular value decomposition $I- \Phi = U\Sigma V^T$ and then replaces $I- \Phi$ with $UV^T$, keeping track of the sign factors in the original little Gram matrix so that some set of vectors $|\phi_i\rangle$ can be recovered for the next round of projections. By this method, for example, we found RICs with 2-quantumness as low as $\approx 7.3$. This made it clear that the $A_4$-RIC could not be the end of the story. At this point we turned to constrained optimization methods in hopes of directly minimizing the quantumness.

\section{Unbiased, Rank-1, Parallel Update}
\label{AskMeAgain}

We began our journey into constrained optimization by trying to preserve as many properties of the SICs as we could. So we looked for real unbiased rank-1 POVMs, with post-measurement states proportional to the POVM elements, which minimize $|| I - \Phi ||_{2}$.

Indeed, assuming the POVM is rank-1 with proportional post-measurement states, we can just as well represent our reference device as an $n \times d$ rectangular matrix $F$, with $d=4$ and $n=\frac{1}{2}d(d+1)=10$. The rows of this matrix are the $d$ dimensional unnormalized vectors $|\phi_{i}\rangle$ whose corresponding POVM elements are $E_{i} = |\phi_{i}\rangle\langle \phi_{i}|$. The demand that the POVM elements sum to the identity amounts to the constraint that the \emph{columns} of this matrix be orthonormal. Thus, under the constraint that $F^T F=I$, or more specifically, $||F^{T}F - I||_{2} = 0$, we want to minimize the 2-quantumness $||I-\Phi||_{2}$.  Recall that $\Phi$ is defined through its inverse. Since we are taking our RIC to be rank-1 with post-measurement states proportional to POVM elements, we have
\begin{equation}
	\Phi^{-1}_{ij} = \frac{|\langle\psi_{i}|\psi_{j}\rangle|^2}{\langle\psi_{j}|\psi_{j}\rangle\hspace{0.2cm}}.
\end{equation}
Another way of thinking about this is to begin with the little Gram matrix of the POVM $g = F F^{T}$ and then graduate to the big Gram matrix $G = g \circ g$, where $\circ$ denotes the Hadamard or entry-wise matrix product~\cite{Horn1994}. If we define a matrix $D$ whose columns are the diagonal entries of $g$,  we can write
\begin{equation}
\Phi = D \circ G^{-1},
\end{equation}
and in particular, if we demand that our POVM is unbiased,  this amounts to $\frac{d}{n}G^{-1}$.
We then find the singular values $s_{i}$  of $I-\Phi$ in order to calculate the Schatten $p$-norm to be minimized. Recall that the $\infty$-norm corresponds to the maximum singular value.

Finally, we impose the condition that the vectors be unbiased, which can be expressed by the constraint $||\vec{g} - \frac{d}{n}\vec{1}||_{2} = 0$, where $\vec{g}$ is the diagonal of $g$ and $\vec{1}$ is the vector of all 1's. Then we are position to perform a minimization of $||I-\Phi||_{p}$ with our two constraint functions, the one imposing the POVM condition $\sum_i E_i = I$ and the other imposing that each element has equal trace. The result of the optimization\footnote{The numerical optimizations in this paper were carried out using both \texttt{python} and \texttt{Mathematica}. The basic python tool employed was the sequential least squares constrained optimizer implemented in the open source library \texttt{scipy}. The Jacobians of the objective function and the constraint functions were automatically differentiated and compiled with \texttt{jax} for speed. On the \texttt{Mathematica} side, we employed \texttt{FindMinimum}, and took advantage of the ability to compile the constraint functions to machine code. } for $p=2$ is given by Eq.~(\ref{LittleSmurf}) in terms of the little Gram matrix $g$, which is a rank-4 projector. As we have seen previously, this specifies the vectors $|\phi_i\rangle$ up to an overall rotation. 

\begin{equation}
g=\left(
\begin{array}{rrrr|rrrrrr}
 \frac{2}{5} & \frac{2}{15} & -\frac{2}{15} & \frac{2}{15} \hspace{0.3cm} 
 & 
 \frac{\sqrt{7}}{15} & -\frac{\sqrt{7}}{15} & -\frac{\sqrt{7}}{15} & -\frac{\sqrt{7}}{15}
   & \frac{\sqrt{7}}{15} & \frac{\sqrt{7}}{15} \\
 \rule{0pt}{15pt} \frac{2}{15} & \frac{2}{5} & \frac{2}{15} & -\frac{2}{15} \hspace{0.3cm} & \frac{\sqrt{7}}{15} & -\frac{\sqrt{7}}{15} & \frac{\sqrt{7}}{15} & \frac{\sqrt{7}}{15} &
   \frac{\sqrt{7}}{15} & \frac{\sqrt{7}}{15} \\
\rule{0pt}{15pt} -\frac{2}{15} & \frac{2}{15} & \frac{2}{5} & \frac{2}{15} \hspace{0.3cm} & \frac{\sqrt{7}}{15} & -\frac{\sqrt{7}}{15} & \frac{\sqrt{7}}{15} & \frac{\sqrt{7}}{15} &
   -\frac{\sqrt{7}}{15} & -\frac{\sqrt{7}}{15} \\
\rule{0pt}{15pt} \frac{2}{15} & -\frac{2}{15} & \frac{2}{15} & \frac{2}{5} \hspace{0.3cm} & \frac{\sqrt{7}}{15} & -\frac{\sqrt{7}}{15} & -\frac{\sqrt{7}}{15} & -\frac{\sqrt{7}}{15}
   & -\frac{\sqrt{7}}{15} & -\frac{\sqrt{7}}{15} \\[2ex] 
\hline
 \frac{\sqrt{7}}{15} & \frac{\sqrt{7}}{15} & \frac{\sqrt{7}}{15} & \frac{\sqrt{7}}{15} \hspace{0.3cm} & \frac{2}{5} & -\frac{1}{15} & -\frac{1}{6} & \frac{1}{6} &
   \frac{1}{6} & -\frac{1}{6} \rule{0pt}{20pt} \\
\rule{0pt}{15pt} -\frac{\sqrt{7}}{15} & -\frac{\sqrt{7}}{15} & -\frac{\sqrt{7}}{15} & -\frac{\sqrt{7}}{15} \hspace{0.3cm} & -\frac{1}{15} & \frac{2}{5} 
& -\frac{1}{6} & \frac{1}{6}
   & \frac{1}{6} & -\frac{1}{6} \\
\rule{0pt}{15pt} -\frac{\sqrt{7}}{15} & \frac{\sqrt{7}}{15} & \frac{\sqrt{7}}{15} & -\frac{\sqrt{7}}{15} \hspace{0.3cm} & -\frac{1}{6} & -\frac{1}{6} & \frac{2}{5} & \frac{1}{15} &
   -\frac{1}{6} & \frac{1}{6} \\
\rule{0pt}{15pt} -\frac{\sqrt{7}}{15} & \frac{\sqrt{7}}{15} & \frac{\sqrt{7}}{15} & -\frac{\sqrt{7}}{15} \hspace{0.3cm} & \frac{1}{6} & \frac{1}{6} & \frac{1}{15} & \frac{2}{5} &
   \frac{1}{6} & -\frac{1}{6} \\
\rule{0pt}{15pt} \frac{\sqrt{7}}{15} & \frac{\sqrt{7}}{15} & -\frac{\sqrt{7}}{15} & -\frac{\sqrt{7}}{15}  \hspace{0.3cm} & \frac{1}{6} & \frac{1}{6} & -\frac{1}{6} & \frac{1}{6} &
   \frac{2}{5} & \frac{1}{15} \\
\rule{0pt}{15pt} \frac{\sqrt{7}}{15} & \frac{\sqrt{7}}{15} & -\frac{\sqrt{7}}{15} & -\frac{\sqrt{7}}{15}\hspace{0.3cm}  & -\frac{1}{6} & -\frac{1}{6} & \frac{1}{6} & -\frac{1}{6} &
   \frac{1}{15} & \frac{2}{5} \\
\end{array}
\right)
\label{LittleSmurf}
\end{equation}

This matrix of algebraic numbers was inferred from floating point results, and indeed, the corresponding POVM is informationally complete since $\det{G}\neq 0$. We will henceforth refer to it as the unbiased 2-RIC. Its 2-quantumness can be calculated exactly to be
\be
\frac{3 \sqrt{2991907}}{784} \approx 6.61879967\dots
\ee
This value agrees with the numerical result up to $10^{-14}$. Given that the 2-quantumness for the non-existent SIC would have been 6, it became clear at this point in our journey that we had entered into fertile territory.

Note that the little Gram matrix splits into two parts. The upper left block represents four vectors which are equiangular among themselves. When rescaled by $\frac{15}{8}$, this $4 \times 4$ block forms a rank-3 projector,
\begin{equation}
\left(
\begin{array}{rrrr}
\frac{3}{4} & \frac{1}{4} & -\frac{1}{4} & \frac{1}{4} \\
\rule{0pt}{14pt} \frac{1}{4} & \frac{3}{4} & \frac{1}{4} & -\frac{1}{4} \\
\rule{0pt}{14pt} \!\!\!\! -\frac{1}{4} & \frac{1}{4} & \frac{3}{4} & \frac{1}{4} \\
\rule{0pt}{14pt} \frac{1}{4} & -\frac{1}{4} & \frac{1}{4} & \frac{3}{4} \\
\end{array}
\right),
\end{equation}
which can be interpreted as a non-informationally complete POVM corresponding to four equiangular lines in 3D. The lower right block represents six vectors which are 2-angular among themselves. Specifically, each of these six vectors makes the same angle with four of the others, and a different angle with one of them up to sign. The eigenvalues of

\begin{equation}
\left(
\begin{array}{rrrrrr}
 \frac{2}{5} & -\frac{1}{15} & -\frac{1}{6} & \frac{1}{6} & \frac{1}{6} & -\frac{1}{6} \\
\rule{0pt}{14pt} \!\!\!\! -\frac{1}{15} & \frac{2}{5} & -\frac{1}{6} & \frac{1}{6} & \frac{1}{6} & -\frac{1}{6} \\
\rule{0pt}{14pt} -\frac{1}{6} & -\frac{1}{6} & \frac{2}{5} & \frac{1}{15} & -\frac{1}{6} & \frac{1}{6} \\
\rule{0pt}{14pt} \frac{1}{6} & \frac{1}{6} & \frac{1}{15} & \frac{2}{5} & \frac{1}{6} & -\frac{1}{6} \\
\rule{0pt}{14pt} \frac{1}{6} & \frac{1}{6} & -\frac{1}{6} & \frac{1}{6} & \frac{2}{5} & \frac{1}{15} \\
\rule{0pt}{14pt} -\frac{1}{6} & -\frac{1}{6} & \frac{1}{6} & -\frac{1}{6} & \frac{1}{15} & \frac{2}{5} \\
\end{array}
\right)
\end{equation}
are $\{1, \frac{7}{15}, \frac{7}{15}, \frac{7}{15}, 0 ,0 \}$. Thus this block does not itself correspond to a POVM in 4D. Finally, considering the upper right and lower left off-diagonal blocks, we can see that the four vectors in the first block make equal angles (up to sign) with respect to the six vectors in the second block. Overall then, we have an unbiased RIC-POVM with four unique inner products.

\begin{figure}
\begin{center}
\includegraphics[width=4.5in]{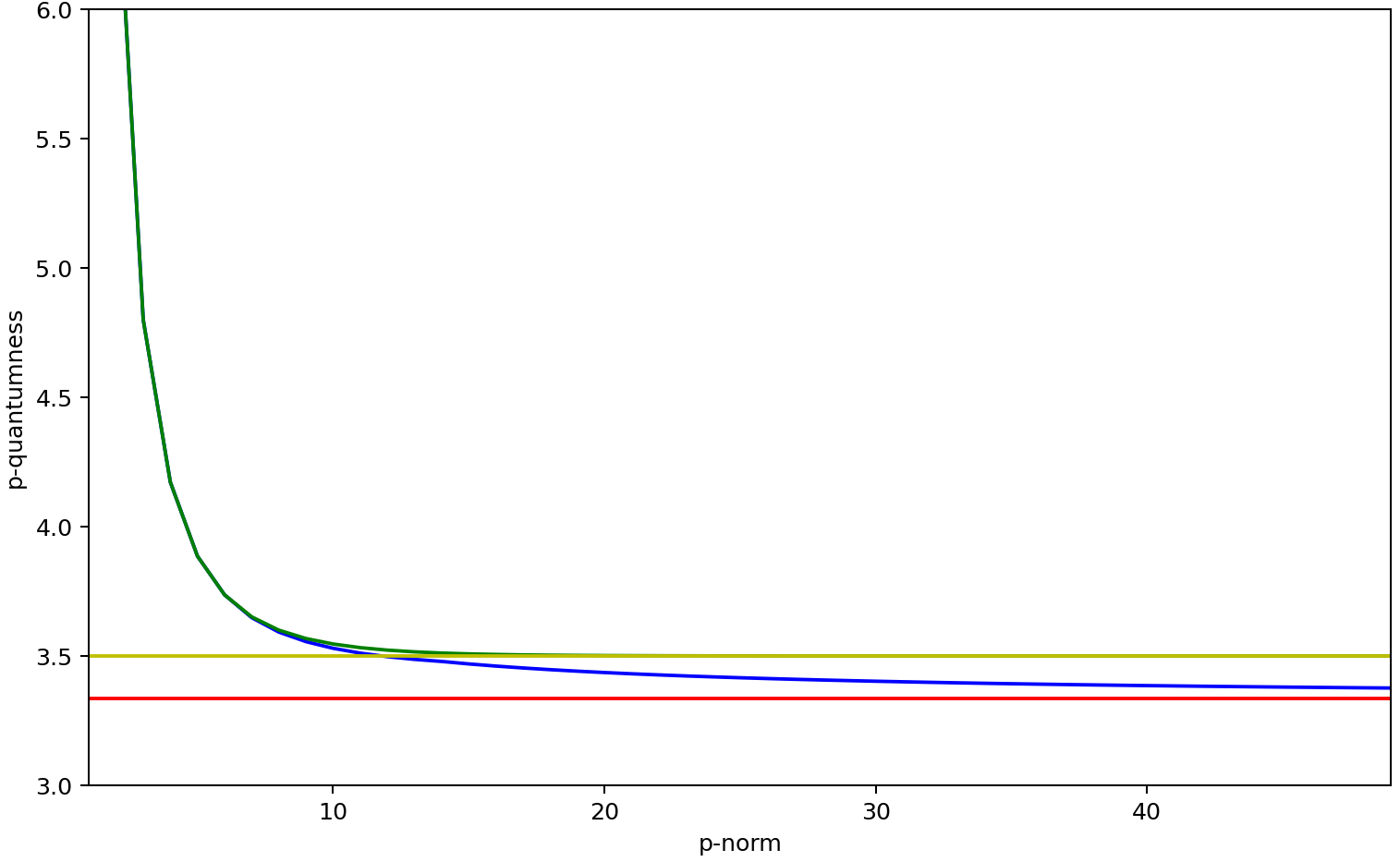}
\end{center}
\caption{The minimized value of the $p$-quantumness for each $p$ over unbiased parallel-update RICs is plotted in blue. The minimized $\infty$-quantumness, to which the former asymptotes, is plotted in red. Meanwhile, plotted in green is the $p$-quantumness of the unbiased 2-RIC: Its $\infty$-quantumness is in yellow. Clearly, the unbiased 2-RIC is not optimal for all values of $p$.}
\label{MattRICQuantumness}
\end{figure}

Of course, we can perform the same optimization for different choices of $p$. These results are displayed in Figure~\ref{MattRICQuantumness}, along with the $p$-quantumness of the unbiased 2-RIC we have been discussing for comparison. There is excellent agreement up to about $p=6$, after which they diverge, the true minima asymptoting to the red line, and the latter to the yellow line. Thus it is clear that the unbiased 2-RIC is \emph{not} univocally a minimally quantum unbiased reference device: different choices of $p$-norm deliver different minima. 

On the other hand, as we've seen, the unbiased 2-RIC has a particularly simple structure, with only five unique entries in its little Gram matrix (up to sign), for which we were able to infer exact algebraic expressions. In contrast, this is \emph{not} true for other unbiased $p$-RIC's. For example, we were unable to find simple algebraic expressions for the little Gram of the unbiased $\infty$-RIC, which appears to have many more than 5 unique entries. Of course, since we are using floating point numbers in our numerical searches, we can only say that a matrix has a certain number of unique entries up to a certain precision. In Table~\ref{DuranDuran} one can see how the number of distinct little Gram entries grows as one considers more decimal places in the case of the unbiased 2-RIC as compared to the unbiased $\infty$-RIC as furnished by our numerical optimization.
\begin{table}
\begin{center}
\begin{tabular}{c|c|c}
Decimal Cutoff & Unique entries (unbiased 2-RIC) & Unique entries (unbiased $\infty$-RIC) \\ \hline 6 & 5 & 9 \\ 11 & 5 & 45 \\ 18 & 55 & 55 \\ 26 & 55 & 55
\end{tabular}
\end{center}
\caption{Number of distinct inner products in little Gram matrix, up to sign.} 
\label{DuranDuran}
\end{table}
Like a SIC, then, the unbiased 2-RIC is at least relatively simple in its structure; in contrast, the unbiased $\infty$-RIC is not. We shall see how this situation changes dramatically when we consider reference devices which are biased.

\section{Biased, Rank-1, Parallel Update}
\label{I'llTellYouTheSame}

Philosophically speaking, it would be a somewhat strange proposal to adopt a reference device with an intrinsic bias, i.e. one for which certain outcomes occur more or less often regardless of the input into the device.\footnote{This point is often emphasized by Blake Stacey.} Nevertheless, in a search for true minimality, to leave no stone unturned, one should consider relaxing the condition that the reference POVM be unbiased. Amazingly, it turns out that one can parameterize a whole family of biased RIC-POVMs by a single variable $f$ delivering a biased RIC which apparently minimizes the $p$-quantumness for any choice of $p$. 

\begin{figure}
\begin{center}
    \includegraphics[width=2in]{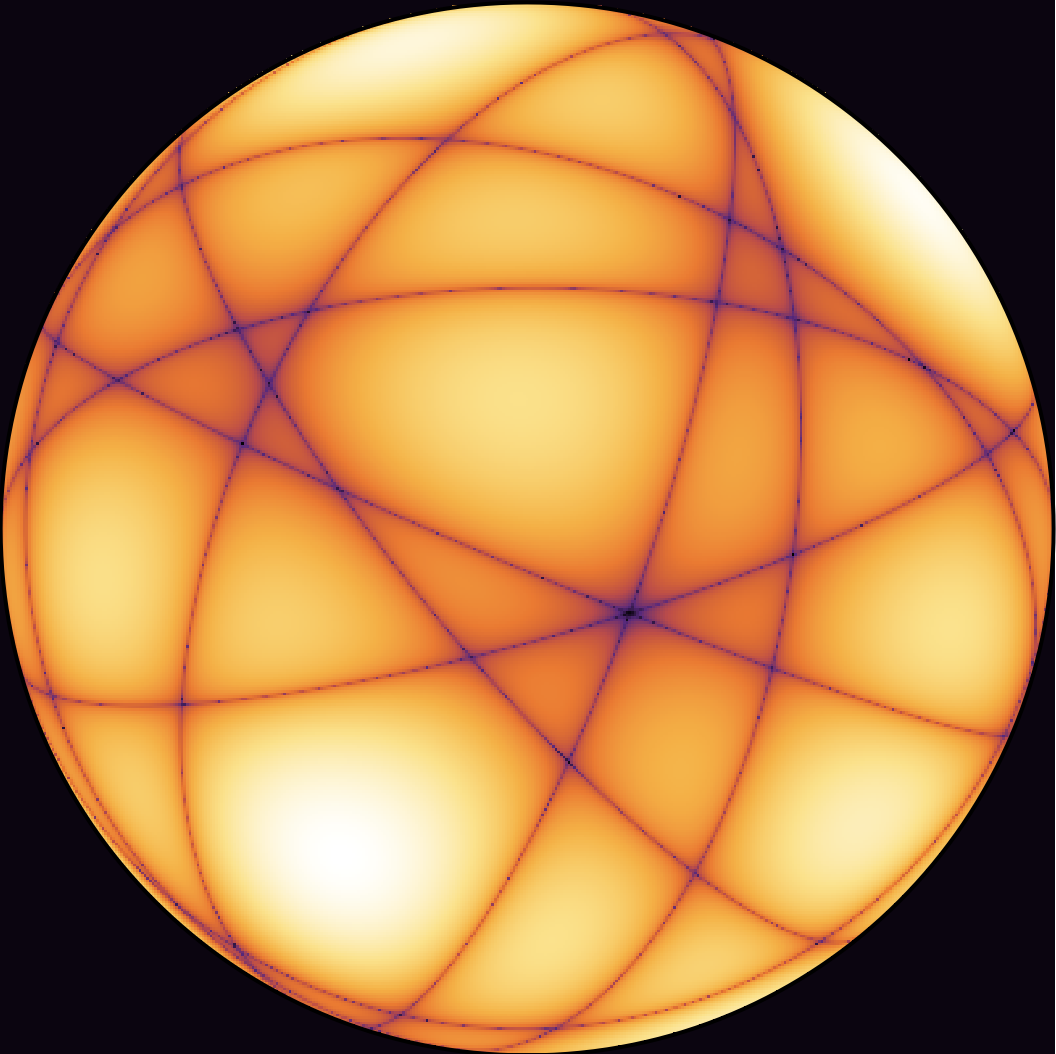}
    \hspace{1cm}\includegraphics[width=2in]{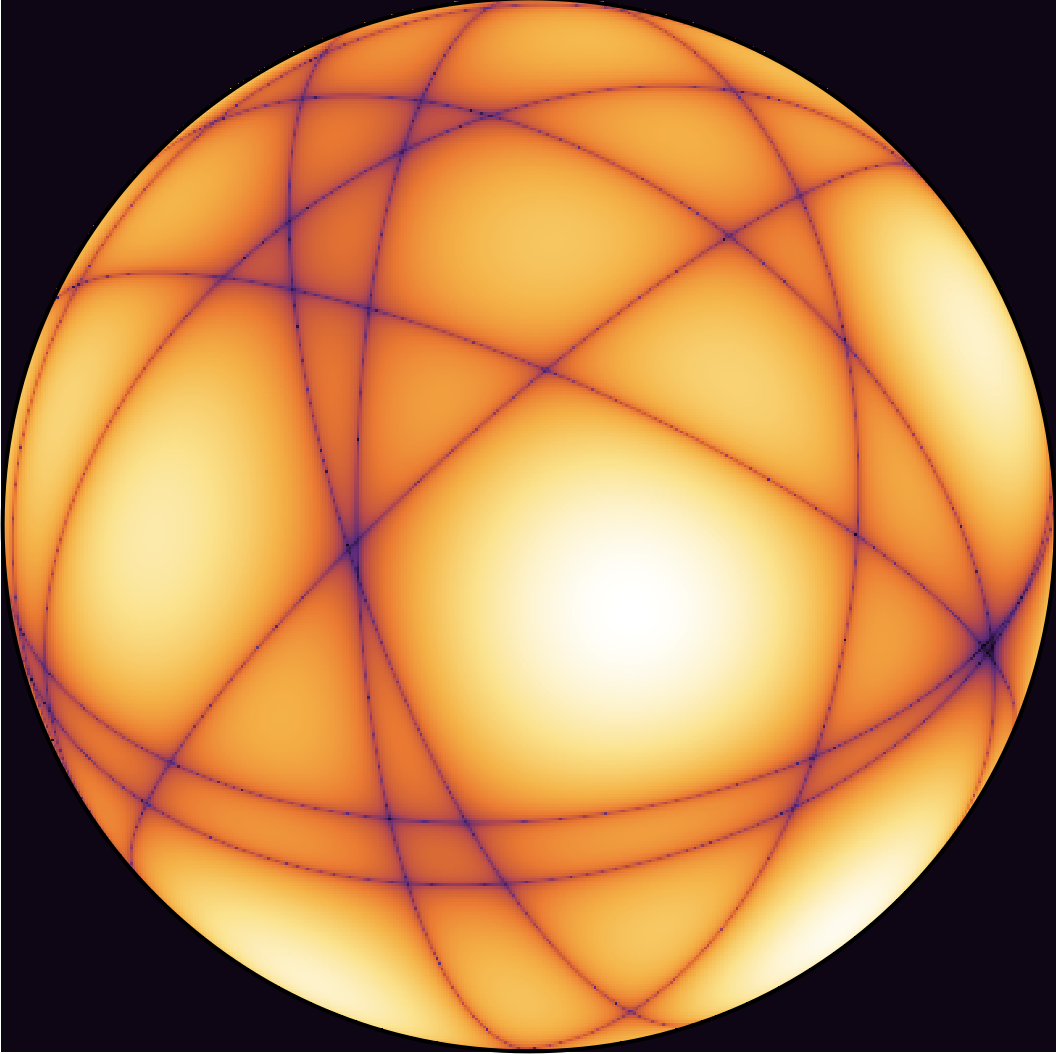}
\end{center}
\caption{On the left, a visualization of the unbiased 2-RIC; and on the right, a visualization of its cousin, the biased 2-RIC.}
\label{MattRICs}
\end{figure}

Indeed, discovering this was a stroke of good fortune. We began by numerically minimizing the 2-quantumness without imposing any constraint on the bias. Inspecting the little Gram matrix of the resulting biased RIC, it became clear that up to sign there were approximately five unique matrix elements. Replacing the numerical values with $5$ unknowns while keeping the sign structure intact (of utmost importance), we obtained the following matrix:
\begin{equation}
g=\left(
\begin{array}{rrrrrrrrrr}
 f & b & -b & b & d & -d & -d & -d & d & d \\
 b & f & b & -b & d & -d & d & d & d & d \\
 -b & b & f & b & d & -d & d & d & -d & -d \\
 b & -b & b & f & d & -d & -d & -d & -d & -d \\
 d & d & d & d & e & -a & -c & c & c & -c \\
 -d & -d & -d & -d & -a & e & -c & c & c & -c \\
 -d & d & d & -d & -c & -c & e & a & -c & c \\
 -d & d & d & -d & c & c & a & e & c & -c \\
 d & d & -d & -d & c & c & -c & c & e & a \\
 d & d & -d & -d & -c & -c & c & -c & a & e \\
\end{array}
\right)\;.
\end{equation}

Next we imposed the rank-1 POVM constraint directly on this matrix, i.e., that the little Gram $g$ must be a rank-4 projector. The $5$ unknowns thus reduced down to a single unknown: $f$. We'll refer to the resulting family of RICs as the \emph{parametric structure}. Interestingly, when $f=\frac{2}{5}$, we recover the unbiased 2-RIC from the previous section.  Indeed, these parameterized RICs have the same structure as the unbiased 2-RIC: five unique entries up to sign, a block of four elements, a block of six elements, with a single angle between them. We call $f$ the bias parameter since it ends up controlling the relative bias between the two blocks: All values $0 < f < \frac{3}{4}$ lead to valid RIC's.

 With this in hand, we can obtain an explicit formula for the $p$-quantumness in terms of the singular values of $I-\Phi$, and thus minimize the parameter $f$ for each choice of $p$. The values of the $p$-quantumness then agree up to $10^{-16}$ or more with those obtained separately through numerical optimization over biased rank-1 parallel-update RIC-POVMs without any assumptions about their structure. So it appears that given any choice of $p$, there is a value of $f$ which delivers a biased RIC which minimizes the $p$-quantumness. In other words, it appears that the minimally quantum biased rank-1 parallel-update RICs are all part of a single parameterized family with a relatively simple structure which takes into account the dependence of the quantumness on the choice of norm. For example, the 2-quantumness finds its minimum at $\approx{6.61544478}$ with $f \approx 0.40446637$: We shall call the resulting RIC the biased 2-RIC.  
 
\begin{figure}
\begin{center}
\includegraphics[width=4.5in]{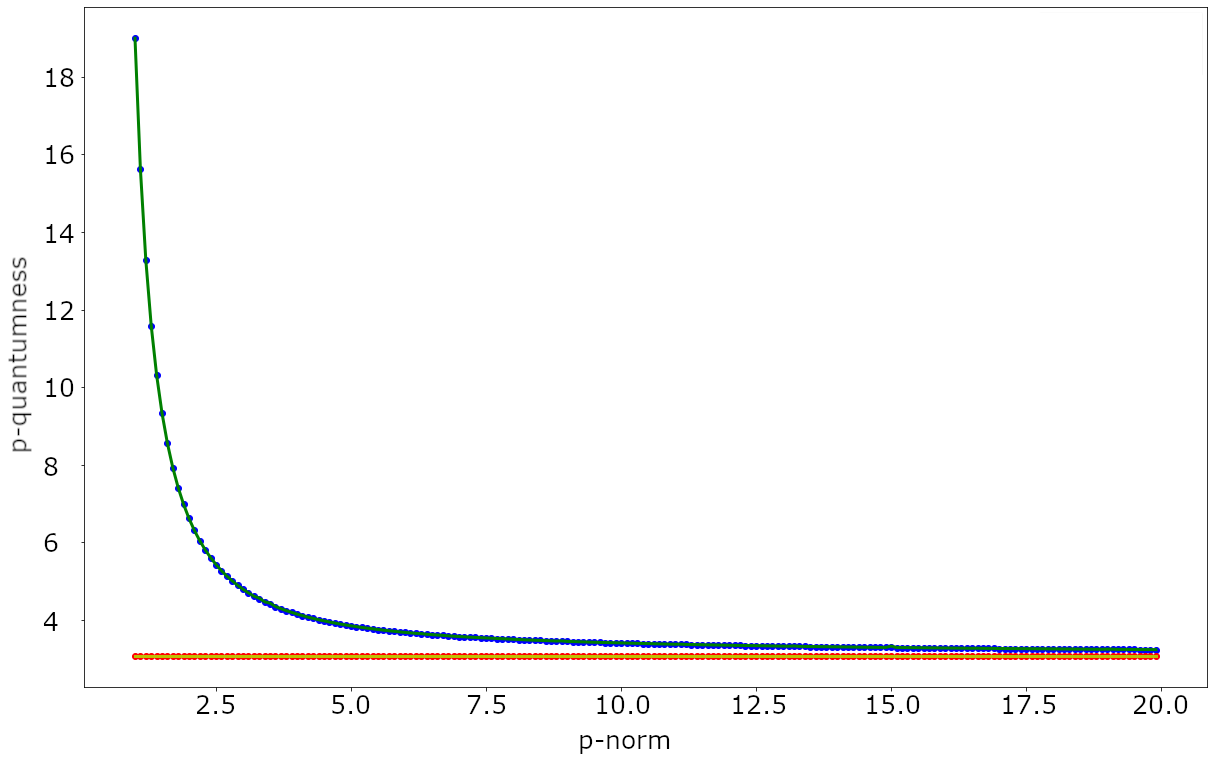}
\end{center}
\caption{The smooth green line depicts the $p$-quantumness minimized over arbitrary rank POVMs (the minimized $\infty$-quantumness is the red line). The blue dots depict the minimized $p$-quantumness over rank-1 POVMs (the minimized $\infty$-quantumness is in yellow). In both cases, the parallel-update rule was used.}
\label{RankN}
\end{figure}

 Furthermore, we performed the same optimization over higher rank POVMs. The rank-1 and arbitrary rank optimizations agree on average up to about $10^{-4}$ (Figure~\ref{RankN}): This is to be expected as the arbitrary rank optimization explores a comparatively more difficult parameter space, essentially that of Kraus operators $K_i$ such that $E_i = K_i^{\dagger}K_i$. Thus it appears the rank-1 assumption is a relatively safe one.
 
Finally, the form of the Born Matrix $\Phi$ in terms of the bias parameter $f$ becomes
\begin{small}
\begin{equation}
\Phi = \left(
\begin{array}{cccc|cccccc}
 q & r & r & r & s & s & s & s & s & s \\
 r & q & r & r & s & s & s & s & s & s \\
 r & r & q & r & s & s & s & s & s & s \\
 r & r & r & q & s & s & s & s & s & s \\
 \hline
 t & t & t & t & u & v & w & w & w & w \\
 t & t & t & t & v & u & w & w & w & w \\
 t & t & t & t & w & w & u & v & w & w \\
 t & t & t & t & w & w & v & u & w & w \\
 t & t & t & t & w & w & w & w & u & v \\
 t & t & t & t & w & w & w & w & v & u \\
\end{array}
\right),
\end{equation}
\end{small}
where
\begin{small}
\begin{align*}
q &= f+\frac{51}{32 f}-\frac{3}{2} && r = f+\frac{15}{32 f}-\frac{3}{2} \\
s &= f-\frac{3}{4} && t = -\frac{(f-1) (4 f-3)}{6 f} \\
u &= -\frac{(f-1)\left(32 f^2-84 f+57\right)}{3 (3-4 f)^2} && v = -\frac{(f-1) \left(32 f^2-12 f+3\right)}{3 (3-4 f)^2}\\
w &= -\frac{4 (f-1) \left(8 f^2-12 f+3\right)}{3 (3-4 f)^2}\;.
\end{align*}
\end{small}
With this, the Born Rule can be written out explicitly in terms of $f$ as
\begin{equation*}
\begin{split}
Q(E_j) = & \sum_{i=1}^{4}\, P(E_j|R_i)
\Bigg[ \frac{9}{8f} P(R_{i}) +
  \left( f+\frac{15}{32f} - \frac{3}{2} \right) \sum_{k=1}^4 P(R_{k}) +
  \left( f-\frac{3}{4} \right) \sum_{l=5}^{10} P(R_{l}) \Bigg] \\[0.75ex]
&
+ \ \frac{1-f}{6f(3-4f)^2} \sum_{i=5}^{10}\, P(E_j|R_i) \Bigg[
\Big( 90f-72f^2 \Big) P(R_{i}) +
\Big( 72f^2-18f \Big) P(R_{\lambda_i})  \\[0.75ex]
&
\hphantom{+ \ \frac{1-f}{6f(3-4f)^2} \sum_{i=5}^{10}\,}
+ \Big( 4f-3 \Big)^{\!3} \sum_{k=1}^4 P(R_{k}) +
\Big( 64f^3 - 96f^2 + 24f \Big) \sum_{l=5}^{10} P(R_{l})
\Bigg]
\end{split}
\end{equation*}
where
\be
\lambda_i = \begin{cases}
i+1 & \text{if} \ i  \ \text{odd} \\
i-1 & \text{if} \ i  \ \text{even}
\end{cases}\;.
\label{HankSnow}
\ee

After meeting this beasty, recall once again what the $\Phi$ for the non-existent real $d=4$ SIC would have looked like:
\begin{equation}
\Phi = \frac{1}{5}\left(
\begin{array}{cccccccccc}
 14 & -1 & -1 & -1 & -1 & -1 & -1 & -1 & -1 & -1 \\
 -1 & 14 & -1 & -1 & -1 & -1 & -1 & -1 & -1 & -1 \\
 -1 & -1 & 14 & -1 & -1 & -1 & -1 & -1 & -1 & -1 \\
 -1 & -1 & -1 & 14 & -1 & -1 & -1 & -1 & -1 & -1 \\
 -1 & -1 & -1 & -1 & 14 & -1 & -1 & -1 & -1 & -1 \\
 -1 & -1 & -1 & -1 & -1 & 14 & -1 & -1 & -1 & -1 \\
 -1 & -1 & -1 & -1 & -1 & -1 & 14 & -1 & -1 & -1 \\
 -1 & -1 & -1 & -1 & -1 & -1 & -1 & 14 & -1 & -1 \\
 -1 & -1 & -1 & -1 & -1 & -1 & -1 & -1 & 14 & -1 \\
 -1 & -1 & -1 & -1 & -1 & -1 & -1 & -1 & -1 & 14 \\
\end{array}
\right),
\end{equation}
which would have given the Born Rule in irreducible QBist form as:
\begin{equation}
Q(E_j)=\sum_{i=1}^{10}\left[3P(R_i)-\frac{1}{5}\right]\! P(E_j|R_i)\;.
\end{equation}
That is really quite some difference.

\section{Non-Parallel Update}
\label{BringingHome}

Finally, we can consider relaxing the parallel-update rule itself. In this case, we must search over not only the RICs, but also the post-measurement states $\sigma_i$. It turns that out we can, in fact, obtain a lower $p$-quantumness than in the parallel-update case. This is true for both the unbiased and biased cases, and what's more, we can obtain a lower value if the post-measurement set itself $\{\sigma_i\}$ is not required to be rescalable to a POVM. That is, it is allowed to be an arbitrary set linearly independent density matrices.

Whereas the equanimity of a SIC implies there is no advantage to having a distinct post-measurement set, evidently the asymmetry of these RICs means that there is. We may note that this is not unlike certain quantum eavesdropping protocols---those in which Alice transmits elements of an ensemble of quantum states to Bob, only for Eve to intercept them first, subject them to a POVM, and on the basis of the outcome, choose from a set of states to send to Bob in hopes of fooling him into thinking she's Alice. One might think the optimal move would be for Eve to pick a post-measurement set proportional to her POVM elements, but this is generally not the case~\cite{Fuchs2000,Fuchs2003}. Unless there is a significant symmetry, the input states, the measurement elements, and the output states will all be different.

In the end, the lowest 2-quantumness we have found so far clocks in at $\approx 6.60798217$. However, we have been unable to find a simple expression for the Born Rule in this case, other than simply pointing to a matrix $\Phi$, full of floating point numbers.

\section{Conclusion}
\label{Lumpitude}

\begin{table}
    \centering
    \begin{tabular}{l|r}
        Candidate & 2-quantumness\\ \hline
        Petersen RIC & 34.0470263 \\
        $A_4$-RIC & 9.1651514\\
        Unbiased 2-RIC & 6.6187997\\
        Biased 2-RIC & 6.6154448\\
        Non-parallel biased 2-RIC & 6.6079822\\
        Non-existent SIC & 6.0000000 \\
    \end{tabular}\\
    \caption{Our results, in short.}
\end{table}

For quantum mechanics over $\mathbb{C}$, SICs provide ideal QBist reference devices. They consist of unbiased, equiangular, rank-1 POVMs with post-measurement states proportional to POVM elements. Moreover, they minimize the quantumness with respect to any unitarily invariant norm. We have seen that for quantum mechanics over $\mathbb{R}$ in $d=4$, the only property to survive is apparently that the POVM and post-measurement set may be rank-1. Not only can lower quantumness be achieved by having biased POVM elements, and by choosing an independent post-measurement set, but even the quantumness itself is no longer a stable quantity: Different reference devices minimize the quantumness with respect to different matrix norms!

It was always clear that the ideal QBist reference device for real-vector-space quantum mechanics must be a more asymmetrical beast, diverging even more from the classical Law of Total Probability than in the complex case, given the lack of a sufficiently large set of equiangular vectors. Our method of exploration has been to proceed by numerical counterexample, and subsequent refinement of the results. It remains to be explained in a positive sense precisely why the somewhat baroque structures detailed in this note must arise. We leave that for a future investigation. Indeed, such an investigation may prove useful beyond quantum foundations, as the structures we've uncovered here may have significance in coding theory (as was the frame that originally inspired this investigation, Ref.~\cite{Bachoc2009,Khanal2020}), or the theory of finite tight frames more generally~\cite{Waldron2018}.

Perhaps the overall message could be summed up in this way. By reformulating quantum mechanics in QBist terms, placing probabilities with respect to reference devices in pride of place, rewriting the Born Rule and even Schr\"odinger's equation in entirely probabilistic terms, one hides what is perhaps one of the most initially striking aspects of quantum theory: its use of complex numbers. Nevertheless, the $i$ is still deeply in the theory: If one drops it and confines oneself to real-vector-space quantum mechanics, its absence is palpable. In the end, the use of complex numbers is really about the symmetry group that underlies quantum theory, one which apparently provides a fertile ground for SICs: Break that and all hell breaks loose as evidenced by the ugliness of the QBist version of the Born Rule. This is why establishing SIC existence in all complex dimensions is such a crucial philosophical issue: If we find that a SIC does not exist in some particular dimension, then the Born Rule when written in irreducible QBist form will likely be every bit as ugly as the expression for $d=4$ real-vector-space quantum mechanics. Indeed, in that case, it would be tempting to regard the probabilistic reformulation as a mere proof-of-principle exercise. In contrast, it is precisely the elegance of the Born Rule in the case of SIC existence, its utterly subtle modification of the Law of Total Probability, that continues to inspire confidence that QBism's philosophical approach is on the right track.

\section*{Acknowledgments}

We thank Hans von Baeyer, John DeBrota, Matthew Graydon, Abhi Khanal, Blake Stacey, and Shayne Waldron for useful discussions and promptings, and we especially thank Blake for his clever suggestion for our title.  This work was supported in part by NSF grant PHY-1912542.

\end{document}